# Neutrino alert systems for Gamma Ray Bursts and Transient astronomical Sources

Stéphane Basa,[a]* Damien Dornic,[b] Gabrielle Lelaizant, [b]

Bruce Gendre,[a] José Busto,[b] Alain Mazure [a]

[a]*Laboratoire d'Astrophysique de Marseille, France*
[b]*Centre de Physique des Particules de Marseille, France*

**Abstract**

GRBs are the most energetic events in the Universe, associated with the death of massive stars (core-collapse supernovae) or the merging of neutron stars or black holes. Discovered in the early 1970s, their cosmological origin was demonstrated only in 1997, when the first distance was measured. Theoretical models predict that the very energetic processes at work in GRBs accelerate charged particles to such energies that they could contribute to the observed high energy neutrinos.
These processes will be discussed and the observational consequences, in particular for current and forthcoming neutrino telescopes, presented.

*PACS*: 95.55.Vj 95.85.Ry 98.70.Rz 98.70.Sa

*Keywords:* neutrinos; Gamma Ray Bursts; Transient Source, detection methods

## 1. The Gamma Ray Bursts

### 1.1. A long story

The history of the Gamma Ray Bursts (GRBs) began in the early 1970s with the observation by the Advanced VELA[1] military satellites of intense and short bursts of gamma-ray radiation (typically between 0.2 and 1.5 MeV). These satellites were especially designed to observe the gamma ray signal produced by possible Soviet nuclear tests in the Earth atmosphere forbidden by the Nuclear Test Ban Treaty. Thanks to their localization capabilities based on the arrival time difference, it was possible to exclude a terrestrial or solar origin. The very first scientific publication reporting the observation of the 16 first GRBs was then published in 1973 [1].

For a very long period, there was a very heated debate in the astronomical community on the origin and the distance of the GRBs: were they produced in our own galaxy or in distant galaxies? Which mechanism could release such an energy?

The BATSE experiment on board the CGRO satellite, by observing more than 2700 GRBs between 1991 and 2000 [2], has shown that these events have remarkable properties:

- They are very frequent: a few per day throughout the Universe;
- Their spatial distribution is isotropic;
- Their duration distribution is clearly bimodal: the 'long' GRBs, for which the duration is higher than two seconds, represent approximately 75% of the events, the others being qualified as 'short' GRBs. This property has been the very first indication that the GRBs could have two different origins.

In February 1997, it was possible to associate for the very first time an X-ray emission to a long GRB, GRB970228, thanks to the Beppo-SAX satellite. Such a measurement was the first from a domain other than gamma rays. After a delay of few hours, a sufficiently accurate position was also achieved to allow the detection of an emission in the visible domain with a large ground-based telescope. This source, termed 'afterglow', decayed over the time-scale of a few days. In May of the same year, another optical afterglow was identified for a long GRB, GRB970508. A spectrum was acquired and the measured redshift was equal to 0.835 [3].
This observation closed a thirty year interrogation: long GRBs occur at cosmological distances. In 2005, this measurement could be also performed for a short GRB, GRB050709. The observed redshift, z=0.16 has shown that short GRBs occur also at cosmological distances. Currently, almost all the GRBs display an X-ray afterglow and, for at least half of them, an emission in the visible domain.

_______
* Corresponding author. Tel.:+33 4 91 05 69 35; fax: +33 4 91 62 11 90; e-mail: Stephane.Basa@oamp.fr.

[1] From the Spanish verb Velar: to Watch.

*1.2. A very promising probe for particle astrophysics*

There is now no doubt that long and short GRBs are produced at cosmological distances, the highest being GRB050904, at z = 6.29 [4]. Taking into account the energy measured on Earth, the typical energy emitted by a GRB, in a couple of minutes, is $10^{51}-10^{52}$ erg, depending on the collimation of the jet, equivalent to the energy emitted by the Sun over the approximate age of the universe or by our entire Milky Way over a few years. GRBs are presently considered as the most energetic events in the universe!

Theoretical models anticipate that the very energetic processes at work in GRBs can make them a promising source of astroparticle phenomena:

- Ultra high energy cosmic rays: AUGER, …
- Gravitational waves: VIRGO, LIGO, …
- Neutrinos: ANTARES, AMANDA, KM3NeT, …

In this paper, we discuss the different possibilities to detect high energy neutrinos produced by a GRB. The different processes which may be at work are discussed in detail in section 2. The consequences on the triggers, and more generally on the observing strategy inherent to specificities of the GRBs, are presented in section 3. Particular interest is addressed to the possibility to extend such a trigger to any transient neutrino source.

## 2. A possible neutrino source

*2.1. The canonical Fireball model*

To explain the GRB phenomena several models have been developed, starting from different hypotheses and describing different scenarios. After one decade of observations, it appears now that the long duration (>2 sec) bursts originate probably from the collapse of a massive stellar progenitor into a black hole, whereas short duration (<2 sec) bursts are from the merger of two compact objects into a black hole. Though these two types of bursts come from two different progenitors, both lead to the formation of a stellar black hole[2] around which an out-flowing plasma is accelerated along jets.

The sudden release of such a large gravitational energy (of the order of a solar rest mass), in a very compact volume (of the order of tens of kilometres), leads to the production of neutrinos, initially in thermal equilibrium with a typical energy of 10-30 MeV, and gravitational waves, not in thermal equilibrium, close to $10^2-10^3$ Hz. These two emissions dominate all the other sources, which are discussed hereafter, even if they are so far undetected!

The leading model to explain the electromagnetic radiation currently observed from the GRBs is not based on the mechanisms occurring during the collapse, but on those taking place in the jets formed during the collapse [4]. In the current canonical picture, the so-called *Fireball* model, the central engine, the newborn stellar black hole, emits "shells" of matter (plasma) into the circumstellar medium with relativistic bulk Lorentz factors (typically Γ>100). If the central engine emits shells with different velocities, a faster shell can reach a slower one and produce a shock. The dissipated energy is used both to accelerate particles and to generate magnetic fields. Charged particles in magnetic fields lose energy via synchrotron emission and, eventually, can boost the synchrotron photons via inverse Compton scattering, producing high-energy photons. These internal shocks are at the origin of the observed *prompt* emission (Fig. 1). When the jets reach the surrounding medium, they start to decelerate, leading to the creation of an external shock. This last one is at the origin of the observed *afterglow* emission.

This model gives a realistic description of the GRB phenomena. However many questions are still opened: which processes generate the energetic ultra-relativistic flow? How is the shock-acceleration realized? …

---

[2] Even if it might form a temporary fast-rotating high-mass neutron star, which collapses to black hole

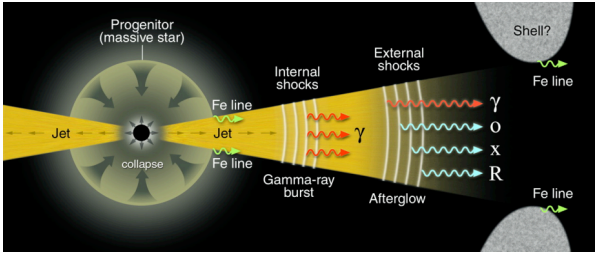

Figure 1: schematic view of a long GRB (gravitational collapse of a very massive star). The internal shocks between the different shells are at the origin of the prompt emission and the external ones of the afterglow emission.

## 2.2. A possible source of ultra high energy neutrinos

### Neutrinos contemporaneous with the gamma rays

In the canonical model, neutrinos may be produced during the creation of the stellar black hole, in the MeV domain, and by the shocks within the jet, in the TeV-PeV domain. We will only consider the last source in this paper. Their energy is more appropriate than those at lower energy - due to their higher interaction cross section - for the neutrino telescopes in operation or in preparation.

In the Fireball model, high-energy neutrinos are expected from photonuclear interactions of the observed gamma-rays produced by synchrotron radiation with the protons accelerated by the internal shocks. Neutrinos and anti-neutrinos are produced in a ratio of 2:1 according to the Delta resonance:

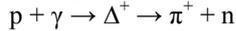

followed by

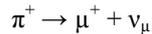

after which the muon will decay to

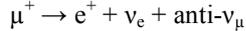

Assuming that the secondary pions receive about 20% of the proton energy per interaction and each secondary lepton shares 25% of the pion energy, each flavor of neutrino is emitted with about 5% of the proton energy, dominantly in the TeV-PeV range. Following the details of the considered model, neutrinos are emitted in coincidence with - or as a precursor (up to several tens of seconds prior) to - gamma-ray emission.

### Neutrinos not contemporaneous with gamma rays

The canonical model of the long GRBs discussed previously is based on the assumption that a relativistic jet is formed. However, at the beginning of the phenomenon the jet is burrowing through the stellar material, and may or may not break through the stellar envelope. Internal shocks in the jets, while still burrowing through the stellar interior, may also produce high energy neutrinos through proton-proton and photomeson interactions [8].

The jets which successfully penetrate through the stellar envelope result in the "standard" GRBs presently observed, and the jets which choke inside the stars, do not produce GRBs. In both cases, neutrinos can pass through the envelope due to their very weak cross section and are emitted as a precursor, about 10-100 s before the neutrinos emitted from the GRB fireball discussed in the previous section. In the case of a choked burst, no electromagnetic signal is detectable; a neutrino signal in isolation may signal a choked GRB.

Conversely, another scenario has been proposed [8], in which a supernova precedes a long-duration GRB by several days to a week. In this supranova model, the supernova remnant provides a natural target for proton-proton interactions, leading to a precursor in neutrinos with energy in the TeV domain.

### The limits of such an exercise

Of course, several predictions have been published. An important limit was computed by Waxman and Bahcall in 1997, who used the measured fluxes of ultra-high energy cosmic rays to renormalize the neutrino fluxes. This assumption leads to a prediction of an annual $\nu_\mu$ flux of $E^2.\Phi_\nu \sim 9\times10^{-9}$ GeV.cm$^{-2}$.s$^{-1}$.sr$^{-1}$, from 100 TeV to 10 PeV (see [6] for a recent calculation). Murase and Nagataki have predicted a similar spectrum for long-duration bursts [7]. However by taking into account the large variability in the GRB parameters, a wider range in the neutrino fluxes is predicted.

Despite all these efforts to predict a neutrino flux, it must be clear that they are, at the end, very uncertain due to all the assumptions inherent in each model. For example, the circumburst environment is not precisely known and the energy of the protons, which has a direct impact on the neutrino flux, is very often arbitrarily renormalized to that of the electrons. Moreover, the inclusion of neutrino oscillations leads to a reduction of these predictions.

In this context, expected fluxes vary considerably from one model to another, and the search for neutrino emission will help to test models of hadronic acceleration in the Fireball model or any other scenario.

## 2.3. Existing limits

The AMANDA detector located in the South Pole has performed the first systematic search for coincidence between the observed photons from a GRB and a neutrino signal [9]. Very recently, stringent limits have been announced by using more than 400 events detected in the Northern Hemisphere. By assuming a Waxman-Bahcall spectrum, a flux upper limit of $E^2 \cdot \Phi_\nu \leq 6.0 \times 10^{-9}$ GeV.cm$^{-2}$.s$^{-1}$.sr$^{-1}$ at 1 PeV, with 90% of the events expected within the energy range of ∼10 TeV to ∼3 PeV, has been announced [10]. Figure 2 shows the expected GRB neutrino flux based on several representative models.

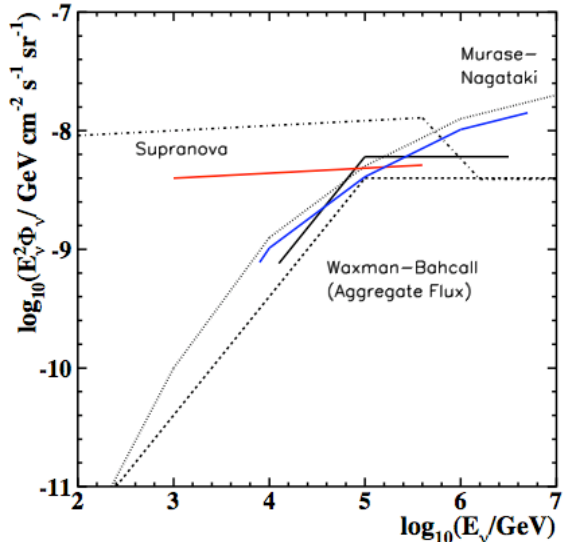

Figure 2: AMANDA neutrino fluxes (solid lines) for muon neutrino energy spectra predicted by the Waxman-Bahcall spectrum (thick-dashed line), the Razzaque et al. spectrum (dot-dashed line) and the Murase-Nagataki spectrum (thin-dotted line) [10]. All fluxes are upper limits and for the entire sky (4π sr).

## 3. The triggers

Two different triggers will be discussed in this section. The first method is the most "natural" one. It is based on the search in conjunction with an accurate timing and positional information provided by an external source: the *Triggered Search*.

The second one exploits the very large field of view of a neutrino telescope. It is based on the search for very high-energy or multiplets of neutrino events coming from the same direction and within the same time window: the *Rolling Search*.

## 3.1. The Triggered Search

It is possible to detect individual bursts in conjunction with an accurate timing and positional information provided by the observation of an event in the gamma-ray domain. In this context, the trigger is delivered by an external source, a satellite. It is then possible to reduce the threshold by using the directional and temporal information. This method has been used by the AMANDA collaboration to compute the limits discussed in the previous section.

The key ingredient in this strategy is the Gamma ray bursts Coordinates Network, (GCN [11]), which distributes freely the localization of the GRBs detected by spacecraft (some in real-time, while the burst is still bursting; others delayed due to telemetry down-link delays). Reports of follow-up observations (the Circulars and the Reports) made by ground-based and space-based optical, radio, X-ray, TeV, and other particle observers are also available.

Today GRB alerts are primarily discovered by the Swift satellite, the dedicated NASA mission which has helped extend the number of measured redshifts, shifting the median redshift from z = 1.0 to z = 2.4 for events prior to, and after, January 2005.

Unfortunately, GLAST - which was launched successfully in June 2008 - will not provide precise positions for the majority of the GRBs that it detects. It will, however, very efficiently complement the energy coverage of SWIFT, which is unfortunately limited at about 150 keV (very often precluding measurement of the peak energy, an important element to characterize the properties of a GRB). The list of the main satellites providing alerts is given in Table 1.

Several GRB missions have been proposed for the next decade. One of the most advanced is the Space–based multi–band astronomical Variable Object Monitor (SVOM), a Sino–French mission dedicated to the detection, localization and study of GRBs and other high–energy transient phenomena. A very special attention is put on the multi-wavelength follow-up of the phenomena (from gamma-ray to X-ray and visible-near infrared). The satellite launch is scheduled in 2012.

Table 1: main satellites providing GRB alerts in 2008.

| Satellite | Instrument | FoV | Localization accuracy | Bands | Operation |
|-----------|-----------|-----|----------------------|-------|-----------|
| INTEGRAL | IBAS | 0.02 sr | 12' | 15 keV – 10 MeV | Oct '02 - |
| SWIFT | BAT | 1.4 sr | 15' | 15 keV - 150 keV | Nov '04 - |
| | XRT | 24' | 18" | 0.2 keV - 10 keV | |
| AGILE | SuperAgile | 1.0 sr | 6' | 15 keV - 60 keV | Jun '07 - |
| GLAST | GBM | 9.5 sr | 1.5° | 8 keV - 30 MeV | Jun '08 - |
| | LAT | 2.4 sr | 15' | 20 MeV - 300 GeV | |

### 3.2. The Rolling Search

It is possible to enhance significantly the sensitivity of a neutrino telescope by using it as a trigger. The basic idea, which has been recently proposed by Kowalski and Mohr [12], is based on the idea that the detection of a burst of neutrinos in temporal and directional coincidences is statistically significant.

For example, with IceCube, the number of atmospheric neutrino doublets is of the order of 10 per year when a temporal window of $\Delta t$=100 sec and a directional window of $\Delta\Omega$=2°×2° is defined. This background becomes negligible, $2.10^{-3}$ $\nu_{atmo}$/year, when a triplet is searched for.

It is also possible to search for single neutrino events, while keeping the background at an acceptable level, by requesting that the reconstructed muon energy is higher than a given threshold energy (typically > 10-100 TeV). A relatively high threshold reduces the atmospheric neutrino background at an acceptable level.

This method offers remarkable potential! There is no hypothesis on the period during which the neutrinos are emitted, a parameter not really constrained in the different models - as discussed in section 2. Moreover only a small fraction of the sky is covered by current gamma-ray, X-ray and optical observatories, whereas neutrino telescopes monitor essentially a half-hemisphere. By using a Rolling Search trigger, it becomes possible to considerably enhance the sensitivity. More importantly, such a method makes no assumption on the nature of the source and the mechanisms occurring inside: it can work on any transient source which produces high energy neutrinos.

### 3.3. The key of the success

The main drawback of the Rolling Search method is that a detection is not automatically associated to an astronomical source. To overcome the problem, it is fundamental to organize a complementary follow-up program. More generally, whatever the method considered, the observations of a transient phenomenon must be organized in order to collect as much information as possible: an *important discovery* needs a complete data set! For example, an observation without a redshift, i.e. without optical spectroscopy, will render almost impossible an estimation of the absolute luminosity of the source.

To organize such a Target-of-Opportunity (ToO) program, it is fundamental to analyze in quasi real-time the neutrino telescope data (< 1 day), transient sources being short-lived phenomena by definition! A good angular resolution is also an advantage, not only to reduce the background, but also to increase the probability to find the counterpart in another wavelength domain. As demonstrated by the Beppo-SAX satellite, a good localization accuracy has made possible the identification of a GRB optical counterpart, while it was impossible with BATSE onboard the CGRO satellite, due to its poor angular resolution (about 1.5°).

With such a philosophy, it will be very important to report a "possible" neutrino observation to the GCN and/or to a follow-up network in order to trigger on complementary observations in case of positive detection.

IceCube in the South Pole and ANTARES in the Mediterranean Sea are organizing such a follow-up program with respectively the RAPTOR (four 45 cm telescopes) and TAROT (two 25 cm telescopes) collaborations. These modest telescopes are perfectly tailored for such a program: since many years, they have been automatically observing the alerts provided by the different GRB satellites.

## 4. Conclusion

ANTARES, IceCube or a forthcoming km$^3$ neutrino telescope will offer a unique opportunity to detect neutrino emission from astrophysical sources. Among all the possible sources, the GRBs, which are the most energetic events since the formation of our Universe, offer very promising perspectives.

The detection of very high energy neutrinos may provide crucial constraints on the fireball parameters and the GRB environment. It would serve as a diagnostic of the presence of relativistic shocks, and as a probe of the acceleration mechanism and the magnetic field strength.

Their detection is simplified by the fact that external triggers are provided by spacecraft currently in operation. A rather modest optical follow-up of interesting events, neutrinos of very high energy or bursts of neutrinos in temporal and directional coincidences (n $\geq$ 2), can significantly improve the sensitivity of the current and forthcoming neutrino telescopes.

More importantly, such a method works on any transient source which produces neutrinos, detected or not by an external trigger. It opens the door to the detection of unexpected phenomena, a very exciting perspective!

## Acknowledgments

S. Basa wishes to thank M. Kowalski for the exciting discussions on neutrino triggers and the organization of the optical follow-up.